\documentclass[manuscript]{aastex}
\usepackage[usenames,dvipsnames,svgnames,table]{xcolor}
\usepackage{amsmath}
\usepackage{natbib}

\shorttitle{}
\shortauthors{Hsieh et al.}

\begin{document}

\title{Molecular Gas Feeding the Circumnuclear Disk of the Galactic Center}

\author{
        Pei-Ying Hsieh\altaffilmark{1},
        Patrick M. Koch\altaffilmark{1},
        Paul T. P. Ho\altaffilmark{1,2},
        Woong-Tae Kim \altaffilmark{3},
        Ya-Wen Tang\altaffilmark{1},
        Hsiang-Hsu Wang \altaffilmark{4},
        Hsi-Wei Yen \altaffilmark{5},
        Chorng-Yuan Hwang \altaffilmark{6}
\\pyhsieh@asiaa.sinica.edu.tw}

\affil{$^1$ Academia Sinica Institute of Astronomy and
       Astrophysics, P.O. Box 23-141, Taipei 10617, Taiwan, R.O.C.}
\affil{$^2$ East Asian Observatory, 660 N. Aohoku Place, University Park, Hilo, Hawaii 96720, U.S.A.}
\affil{$^3$ Department of Physics \& Astronomy, Seoul National University, Seoul 151-742, Korea}
\affil{$^4$ Department of Physics and Institute of Theoretical Physics, The Chinese University of Hong Kong, Shatin, New Territories, Hong Kong, China}
\affil{$^5$ European Southern Observatory (ESO), Karl-Schwarzschild-Str. 2, D-85748 Garching, Germany.}
\affil{$^6$ Institute of Astronomy, National Central University, No.300, Jhongda Rd., Jhongli City, Taoyuan County 32001, Taiwan, R.O.C.}

\begin{abstract}
The interaction between a supermassive black hole (SMBH) and the surrounding material is of primary importance in modern astrophysics. The detection of the molecular 2-pc circumnuclear disk (CND) immediately around the Milky Way SMBH, SgrA*, provides an unique opportunity to study SMBH accretion at sub–parsec scales. Our new wide-field CS($J=2-1$) map toward the Galactic center (GC) reveals multiple dense molecular streamers originated from the ambient clouds 20-pc further out, and connecting to the central 2 parsecs of the CND. These dense gas streamers appear to carry gas directly toward the nuclear region and might be captured by the central potential. Our phase-plot analysis indicates that these streamers show a signature of rotation and inward radial motion with progressively higher velocities as the gas approaches the CND and finally ends up co-rotating with the CND. Our results might suggest a possible mechanism of gas feeding the CND from 20 pc around 2 pc in the GC. In this paper, we discuss the morphology and the kinematics of these streamers. As the nearest observable Galactic nucleus, this feeding process may have implications for understanding the processes in extragalactic nuclei.
\end{abstract}


\section{INTRODUCTION}\label{sect-intro}

\begin{figure}
\begin{center}
\epsscale{0.5}
%/sma/kea/pyhsieh/smadata/GalacticCenter/inflow
\centering
\makebox[0pt]{\includegraphics[angle=0,scale=0.4]{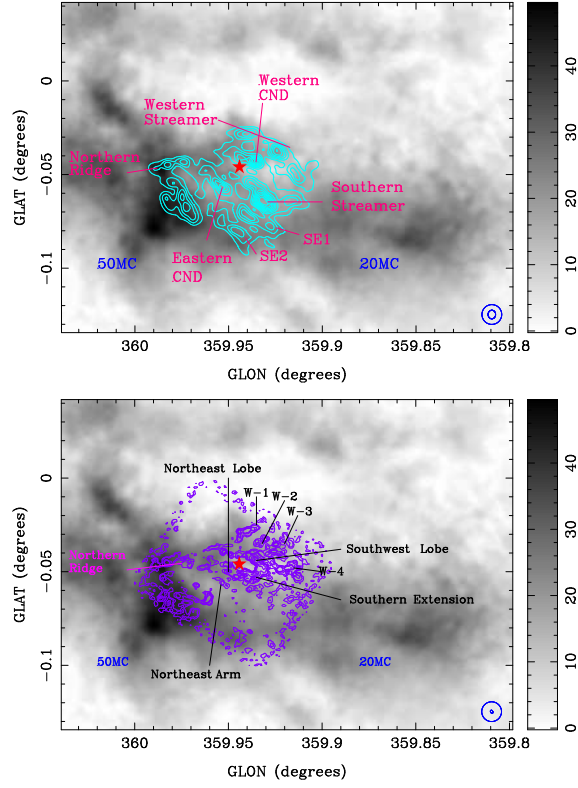}}
\caption[]{Upper: Integrated intensity map (MOM0) of our new CS($J=2-1$) data (grey) \citep{hsieh16}. The VLA NH$_{3}$(3,3) MOM0 map \citep{herrnstein02} is overlaid (contours in 10\%, 20\%, ..., 100\% of the peak, where the peak is 23.2 Jy beam$^{-1}$ km s$^{-1}$). The resolutions of the CS($J=2-1$) and NH$_{3}$(3,3) are 20$\arcsec$ and 16.6$\arcsec\times14.5\arcsec$, respectively.
We labeled the 20 MC, 50 MC, western and eastern CND, and the streamers traced by NH$_{3}$(3,3). The streamers \citep{mcgary01} include the western streamer, northern ridge, southern streamer, and the SE1/SE2. SgrA* is labeled with a red star. Lower: HCN($J=4-3$) contours overlaid on our CS map. The beam size of the HCN data is $5.87\arcsec \times 4.44\arcsec$. The contour levels are, 2, 4, 8, 10, 20, 40, 60, 80, 100$\times$10 Jy beam$^{-1}$ km s$^{-1}$. The HCN($J=4-3$) CND consists of features called the southwest lobe, the southern extension, the northeast arm, and the northeast lobe \citep{chris,maria,liu12}. Several protrusions called W-1, W-2, W-3, and W-4 are also labeled \citep{liu12}. W-1, W-2, and W-3 are the counterparts of the western streamer seen in the NH$_{3}$(3,3) line.
}
\label{fig-mom0}
\end{center}
\end{figure}

The origin of the 2-pc CND in the GC has remained unclear in spite of intensive study for the past decades \citep[e.g.,][]{guesten87,jackson93,amo11,harris,mezger,etx,lau,wright,maria,martin12,herrnstein02,
herrnstein05,chris,great,mills13b}.
The CND is a ring-like molecular structure rotating with respect to the supermassive black hole SgrA*, within which are the arc-shape ionized gas streamers called SgrA West \citep{roberts93}. The SgrA West arms converge at SgrA* and have been proposed to originate from the inner edge of the CND. The CND, being the closest molecular reservoir in the GC, is critical on the understanding of the feeding of the nucleus. The replenishment of the CND itself, therefore, is an important problem.
Thus, many observations have been made to detect the kinematic connections between nearby molecular clouds and the CND with the main purpose of detecting the inflow of gas to the CND. 
Previous NH$_{3}$(3,3) observations carried out with the VLA have detected several "streamers" \citep{okumura89,ho91,coil00,mcgary01} (Figure~\ref{fig-mom0}).
The 20 km s$^{-1}$ cloud (hereafter 20 MC) and the 50 km s$^{-1}$ cloud (hereafter 50 MC) lying 20 pc south of SgrA$^{*}$, appear to morphologically connect to the CND, with several NH$_{3}$(3,3) streamers called the ``southern streamer'', ``SE1 streamer'', and the ``northern ridge''. The southern streamer extends northward from the 20 MC toward the southeastern edge of the CND \citep{okumura89,ho91,coil00}. The northern ridge originated from the inner edge of the 50 MC is also connected to the CND with a velocity gradient of 0.5 km s$^{-1}$ arcsec$^{-1}$ spanning 110$\arcsec$ (4 pc). SE1 extends northward and kinematically connects to the eastern lobe of the CND.

The northern ridge, southern streamer, and SE1 were proposed to feed the CND based on observed increasing line widths and heating as they approach the GC.
However, the understanding of the accretion of the gas and that of the CND were limited by the previous lower-transitions and lower-resolution data \citep[e.g.][]{tsuboi99}. For instance, the CND and the region interiors were not well sampled by the NH$_{3}$(3,3) line, but are prominent in the NH$_{3}$(6,6), HCN and CS lines \citep[e.g.][also see Figure~\ref{fig-mom0}]{maria}. This suggests that from the CND inward, the gas is dense and hot. The streamers also become faint toward the CND, and this suggests that the HCN and CS dense gas tracers are more reliable to detect the connecting material.
Previous interferometric data also suffered from the negative bowel effect due to abundant missing information of the large-scale structures. Thus, single-dish mapping is essential to recover the large-scale component. Moreover, wide-field mapping is also important to trace the origin of the streamers from the ambient clouds (20/50 MC). Therefore, we observed the central 30 pc with the Nobeyama Radio Observatory (NRO) 45-m and the Caltech Submillimeter Observatory (CSO) with the CS $J_{u}=5,4,2$ lines. The details of the observations are described in \citet{hsieh15,hsieh16}, where we presented the studies of the {\it molecular outflow} in the central 30 pc of the GC with 40$\arcsec$ resolution maps. In this paper, we focus on the investigation of the {\it molecular inflow} associated with the CND utilizing the higher resolution of the 20$\arcsec$ in the CS($J=2-1$) map.

\section{Comparing  NRO 45-m/CSO and  SMA data -- Revisiting  Structures}

In Figure~\ref{fig-mom0} and Figure~\ref{fig-chan} we show the integrated intensity and the channel maps of our NRO-45m CS($J=2-1$), CSO CS($J=5-4$) and the SMA HCN($J=4-3$) data \citep{maria,liu12}, respectively.
While the channel maps in \citet{hsieh16} are shown with a low-velocity resolution of 40 km s$^{-1}$, we display maps here with 5 km s$^{-1}$ resolution.
Here, we note the well-studied components in the central 30 pc of the GC. In the  HCN($J=4-3$) integrated intensity map (beam$=5.87\arcsec\times4.44\arcsec$), the CND appears as a ring-like structure. The western streamers outside the CND are resolved into protrusions called W-1 to W-3 in the SMA map.
The southern lobe and the northern ridge are also present in the HCN($J=4-3$) map. The W-4 component is proposed to connect to the southern lobe in \citet{liu12}. This is also called the negative-longitude extension (NLE) by \citet{oka11} and \citet{takekawa17}.
The HCN($J=4-3$) emission of the southern streamer is fainter than its NH$_{3}$(3,3) emission. The 20/50 MC located south of the CND are clearly seen in the CS maps.

A comparison between the CS and HCN($J=4-3$) channel maps reveals that there is a streamer in CS, a counterpart to the western streamer, which is visually connected to the 20 MC through a linear feature around $-18$ km s$^{-1}$ to 2.5 km s$^{-1}$ (bounded by two cyan lines in Figure~\ref{fig-chan}).  
We call this linear feature ``extended western streamer (extW-streamer)'' in this paper.
The extW-streamer can also be seen in the previous dense gas maps \citep{tsuboi99,amo11,liu12}, but it is not mentioned before due to its lower brightness in several $J=1-0$ lines. Our new CS maps are more sensitive to both diffuse and compact components.

In Figure~\ref{fig-west}, we present the CS($J=2-1$) map integrated from $-24$ km s$^{-1}$ to $-11$ km s$^{-1}$ (grey) overlaid on the integrated HCN($J=4-3$) map. The velocity ranges are selected to avoid the contamination of the 20 MC. The extW-streamer appears as a smooth ridge-like feature visually connecting the 20 MC and the western streamer. The resolved W-1 to W-3  protrusions appear as ''compact cores'' within the extW-streamer near the CND. The spectra of the extW-streamer are shown in Figure~\ref{fig-west}. The SMA spectra are smoothed to the same resolution (20$\arcsec$) as the NRO 45-m data for comparison. The extW-streamer shows narrow line widths (full width half maximum (FWHM) of 30 km s$^{-1}$) at the positions 7, 8, 9 and becomes broader as it becomes the western streamer near the CND (FWHM of 80 km s$^{-1}$). Note that at the positions 7, 8 and 9, a feature called connecting ridge (CR) appears at the velocity around 40 km s$^{-1}$ as discussed in \citet{hsieh15}. The CR is proposed to be elevated disk emission connecting to the extraplanar feature called the polar arc (PA). This CR appears to have no physical association with the CND nor the streamers.
The CS($J=2-1)$ spectra show a wind component in the western streamer, which is absent in the HCN($J=4-3$) spectrum. This wind component might be filtered out in the interferometric data if the broad wind component is extended.

{\bf The NRO 45-m CS(2-1) data can be found on the following link.}
\url{https://drive.google.com/open?id=0B87dHJ1XIqjwSVpEdXFKVm01cVU}

\section{Large-Scale Streamers}

\subsection{Western/extW and Eastern Streamer}

{\it\underline{extW and Western Streamer} -} In Figure~\ref{fig-pv-west} we show the CS($J=2-1$) and CS($J=5-4$) longitude-velocity diagrams (lv-diagrams) sliced along the extW-streamer from latitude $(b)=-0.035\degr~\rm to-0.022\degr$. In the lv-diagrams, there are two streamers running nearly parallel with a velocity difference of $\sim$10 km s$^{-1}$. These two components show different kinematic trends. One of the streamers (green and yellow squares) increases in velocities progressively as it approaches and stops near the CND at the velocity of $\sim$100 km s$^{-1}$. This streamer is the ext-W streamer (squares) and the western streamer (yellow and red squares in the upper and lower panels, respectively). The other streamer remains at a constant velocity around $-16$ km s$^{-1}$ (yellow and red lines in the upper and lower panels, respectively). The line widths of the extW and the Western streamer are also increasing as they approach the CND by about a factor of two. The $-16$ km s$^{-1}$-streamer, on the other hand, maintains the same line width.
The lv-diagrams indicate that the extW-streamer originates from the ambient cloud and moves towards the CND. Our wide-field map reveals this structural connection better than previous maps.
Moreover, the extW-streamer also exists in CS($J=5-4$), while the $-16$ km s$^{-1}$ streamer disappears. This suggests a high-excitation nature of the extW-streamer, consistent with being physically located in the GC.

{\it\underline{Eastern streamer} -} Figure~\ref{fig-pv-east} displays CS($J=2-1$) and CS($J=5-4$) lv-diagrams sliced along the eastern part of the CND (the sense of ``eastern'' is in the conventional celestial coordinate relative to the western streamer). We find that there is a high velocity-component from 60 km s$^{-1}$ to $-100$ km s$^{-1}$ in Figure~\ref{fig-pv-east} from $(l, b)=(360\degr, -0.067\degr)$ to $(359.9\degr, -0.055\degr)$. The spectra in Figure~\ref{fig-east} show two velocity components. The higher velocity component traces the southern lobe of the CND and connects to the protrusions east of the CND in the celestial plane. This component shows a kinematic behavior similar to the western streamer. We call this feature ``eastern streamer''.
The eastern streamer is also seen in CO($J=3-2$) \citep{oka11} but it is affected by  foreground absorption.
Significantly more extended and connected features are seen in the CS($J=2-1$) map. In Figure~\ref{fig-east}, the eastern streamer seems to connect to the northern ridge, which is a linear feature located at a velocity of $\sim-10$ km s$^{-1}$. \citet{mcgary01} reported that the northern ridge shows a kinematic connection to the CND.
However, in Figure~\ref{fig-pv-east}, the origin of the eastern streamer might be from the 50 MC. A part of the 50 MC shows elongation in the lv-diagrams from $b=-0.067\degr$ and seems to be stretched towards the eastern streamer. The physical associations of the 50 MC and the northern ridge are unclear because of the interaction with the supernova remnant SgrA East \citep{serabyn92,lee03}.
An additional component, called C1 by \citet{oka11}, shows a kinematic behavior similar to the eastern streamer.

%Similar feature is seen in \citet{oka11} with the high CO($J=3-2$)/CO($J=1-0$) components.

Note that the eastern streamer is different from the southern streamer.
The southern streamer has velocities of $\sim~20-40$ km s$^{-1}$. This suggests that the southern streamer is superimposed on the eastern streamer along the line of sight.
The kinematics of the southern streamer show no significant velocity gradients and do not appear to be directly associated with neither the CND nor the nucleus \citep{nh3_05}, unless the southern streamer is moving perpendicular to the line of sight.

\subsection{Longitude-Velocity Diagrams}

In Figure~\ref{fig-pv-two-ring}, we compare the extW-streamer, the eastern streamer, and the CND in lv-diagrams in HCN($J=4-3$) and in CS($J=2-1$). The extW-streamer is averaged from $b=-0.035\degr$ to $b=-0.022\degr$. The eastern streamer is averaged from $b=-0.067\degr$ to $b=-0.055\degr$, and the CND is averaged from $b=-0.053\degr$ to $b=-0.04\degr$.
The majority of the extW-streamer (red contours) is moving with positive velocity and the eastern streamer (blue contours) is moving with negative velocity.
These two streamers show rotating patterns from $(l,v)=(-0.02\degr, 80~\rm km~s^{-1})$ to $(l,v)=(0.02\degr, -80~\rm~km s^{-1})$.
The CND is rotating from $(l,v)=(-0.01\degr, 125~\rm km~s^{-1})$ to $(l,v)=(-0.01\degr,-120~\rm km~s^{-1})$.  Thus, the extW/Southern streamers can be described with a rotation that is slower than the CND's. Yet another rotating system are the 20/50 MCs that show large-scale rotation of the Galactic disk. These results hint the co-existence of multiple rotating systems in the central 12 pc of the GC. Since HCN($J=4-3$) has a higher excitation than CS($J=2-1$), it samples the CND better. On the other hand, CS($J=2-1$) traces extended emission/low excitation structures, i.e., the extW-/eastern streamers. Our results provide further evidence that the streamers have lower excitation and are located outside of the CND.
Similar lv-diagrams of the ext-W/eastern streamers are presented by \citet{oka11} in CO data. 
In their interpretation, the extW/eastern streamers are the outer part of the CND, which is still infalling to the CND with an infall velocity of 50 km s$^{-1}$.
However, the CO data can be contaminated by foreground absorption, and \citet{oka11}
extracted the highly excited gas with CO($J=3-2$)/CO($J=1-0$) ratios $\ge1.5$. This extraction relying on intensity ratios can be uncertain and arbitrary.
Since our dense gas map traced by CS($J=2-1$) is less affected by foreground emission, we can more cleanly depict the kinematic structures.
Nevertheless, if the high-velocity ``ring'' is physically located in the GC, the lv-diagrams are not suitable for comparison because one fixed l corresponds to different radii of the ring. Therefore, in order to clarify the kinematics of the connection between the streamers and the ambient clouds, we show position-velocity diagrams  drawn on trajectories along the ``ring'' (phase plot) \citep[e.g.][]{jackson93,martin12} in the next section.

\section{Phase Plots: Kinematics of the CND, the Eastern and extW-Streamers}\label{sect-phase}

In Figure~\ref{fig-phase-st1} and Figure~\ref{fig-phase-st2} we present the phase plots of the Eastern/extW-streamers  drawn along a projected circle (red elliptical annulus) centered on SgrA* in CS($J=2-1$) and HCN($J=4-3$).
The intrinsic radii of ellipse (circle), inclination angle, and position angles ($PA$;  angle of the (geometric) major axis of an ellipse in the plane of the sky measured from the East) of the ellipse is 5 pc (4 pc), 65$\degr$, and  $-4\degr, 6\degr, 16\degr$, respectively. The $PA$ is defined starting from east clockwise. The inclination angle and $PA$ of $6\degr$  (geometric major axis of the CND) are the adopted from \citet{jackson93}.
The HCN($J=4-3$) data are smoothed to 20$\arcsec$ to match the CS($J=2-1$) data. In the HCN($J=4-3$) map, the ``streamers'' are fluffy structures surrounding the CND and mix with the 50/20 MC along the line of sight. The elliptical annulus and the starting angle (intersection of the vertical/horizontal black lines) are labeled in the intensity maps. The direction of the phase plot is counterclockwise along the elliptical annulus. Our goal here is to investigate whether these rings/streamers are gravitationally bound by the central mass concentration (i.e., SgrA* and stars) and whether there are co-rotating systems.

Relative to the lower velocity components of the 20/50 MC, high-velocity components are seen in both CS and HCN with maximum velocities of $\sim~-120$ km s$^{-1}$ and $\sim~120$ km s$^{-1}$, which correspond to the eastern and the extW-streamers, respectively. The overall kinematic structures in CS($J=2-1$) and HCN($J=4-3$) are consistent, but HCN($J=4-3$) recovers more compact components in the 20/50 MC and the extW/eastern streamers. The western streamers identified in the interferometric NH$_{3}$(3,3) and the HCN($J=4-3$) maps show a CS($J=2-1$) counterpart in the NRO 45-m data (the extW-streamer).
For a more quantitative model comparison, we generate Keplerian rotation curves with a simplified assumption of circular orbit and we overlap these on the high-velocity components in the phase plots.
To further distinguish different scenarios, an additional infall velocity ($V_{\rm infall}$) is implemented. The resulting projected velocity ($V_{\rm rad}$) measured along the line of sight then becomes.
\begin{equation*}\label{eq1}
V_{\rm rad} = \sqrt{\frac{GM(r)}{r}}\sin(i)\cos({\rm \theta-PA})+V_{\rm infall}\sin(i)\sin({\rm \theta-PA}),
\end{equation*}
where $i$, $r$, and $PA$ are the inclination, radius, and position angle of the major axis, respectively, while $\theta$ denotes the azimuthal angle measured counter-clockwise from the $PA$.
The enclosed mass $M(r)$ is a function of radius including the masses of star and SgrA* \citep[$M(\rm SgrA*)=4\times10^{6}~M_{\odot}$;][]{ghez05}.
In the phase plots, we overplot $V_{\rm rad}$ by taking $r=5$ or 4 pc, $i=65^{\circ}$, and $PA=-4\degr, 6\degr, 16\degr$, for $V_{\rm infall}=0$, 20 km s$^{-1}$, 50 km s$^{-1}$ and 70 km s$^{-1}$. The enclosed mass $M(\rm r = 4\rm~pc)$ and $M(\rm r = 5\rm~pc)$ are $9.9\times10^{6}~M_{\odot}$ and $11.8\times10^{6}~M_{\odot}$ \citep[see equation 5 in][]{genzel10}, respectively.
We find that the gross features of the extW/eastern streamers in both lines can not be described by a pure Keplerian rotation, where the max/minimum values at $\theta=0~{\rm or}~180\degr$ if the gas were subject to pure rotation.
In fact, we find an infall motion ($V_{\rm infall}$ = 20-70 km s$^{-1}$) is needed to explain the gross features of the streamers.
The loci of the max/minimum $V_{\rm rad}$ are at $\theta=-20~{\rm and}~160\degr$ indicates that there should be non-negligible radial gas motions, which we found amount to 20-70 km/s. This range in infall motion defines a band in the phase plot that appears to capture the extended emission of the streamers.

In Figure~\ref{fig-phase-cnd1} and Figure~\ref{fig-phase-cnd2} we show the phase plots of the CND along a projected circle with radius of $r=3.5$ pc for the upper panels and $r=2.5$ pc for the lower panels. The Keplerian curves are overlaid on the phase plots of the CND with $r=$3.5 pc (2.5 pc in the lower panels), $i=65\degr$, and $PA=-4\degr, 6\degr, 16\degr$ for $V_{\rm infall}=$0, 20, 50, and 70 km s$^{-1}$.
The CND shows a kinematic behavior similar to the streamers, which indicates that they are  a co-rotating system. With a detailed inspection, we further find that the western streamers seem to intersect with the CND with increasing velocities. Besides, the phase-shifting of the western streamers toward the CND may also suggest variations in the curvatures of the western streamers.
The $V_{\rm infall}=20\sim70$ km s$^{-1}$ is still needed to describe the kinematic feature of the western streamer, but the S-extension and the NE-lobe of the CND shows pure Keplerian motion at radius of 2.5 pc, which is consistent to the previous results \citep[e.g.][]{jackson93,martin12}.
Different from the $lv$-diagrams (Figure~\ref{fig-pv-two-ring}), the velocity difference of the streamer and the CND is more subtle in the phase plots, where the streamers and the CND show similar rotation patterns with the same velocity of approximately  $\pm120$ km s$^{-1}$. The high velocity components of $\pm120$ km s$^{-1}$ (the W-4 and the northeast arm) are not included in the streamers.
Our results confirm that the conventional $lv$-diagrams are not suitable for studying the ring-systems because different $l$ corresponds to different radii of the ring.

The association of the eastern streamer and the southern extension of the CND is unclear, but the southern extension shows a faster rotation than the eastern streamer.
The eastern streamer can possibly be a low-excitation component of the southern extension of the CND. The southern extension seems to be better described by a pure Keplerian rotation at a radius of 2.5 pc, which is consistent with the results of \citet{jackson93,martin12}.
The W-4 component also connects to the southwest lobe of the CND. A more sophisticated modeling beyond the scope of this paper will be required to fit the data by combing interferometric and single-dish maps together.

\section{Discussion}

\subsection{On-Going Accretion onto the CND}

Our joint analysis of the NRO 45-m and the SMA data shows that the extW- and the eastern streamers consist of both compact and extended structures.
With the moderately high-J CS line ($J=2-1$), we find that the eastern streamer might be originated from the 50 MC (Figure~\ref{fig-pv-east}). The stretching of substructures in the 50 MC suggests that the gas is being tidally disrupted and dragged to the inner orbit with progressively higher velocities up to $-120$ km s$^{-1}$ (projected). Based on our simple model in Sect.~\ref{sect-phase}, at a distance of 4.5 pc from SgrA*,  with a moving velocity of $\sim$130 km s$^{-1}$ (intrinsic), the rotation time of the eastern streamer is 7$\times10^{4}$ years to cover 1/3 of the orbit.
Different from the eastern streamer, we do not see clear evidence that the extW-streamer is connected to the 20/50 MC. However, the kinematic connection of the eastern streamer, the W-4 component, and the extW/western streamers shown in the phase plots may suggest that these features are on the same circular orbit. In this scenario, these surrounding streamers may originate from the 50 MC and are formed more than $7\times10^{4}$ years ago.
An alternative model to explain the formation of the W-4 component and the western streamer is proposed by \citet{takekawa17}. These authors proposed that the W-4 component is smoothly connected to the 20 MC through the ``bridge'' with velocities from $-20$ km s$^{-1}$ to 30 km s$^{-1}$ at $(l,b)=(359.91\degr,-0.05\degr)$ (see Figure 3 in their paper). We locate the bridge in Figure~\ref{fig-phase-st1} and ~\ref{fig-phase-st2}. Even though there is a visual connection between the bridge and the W-4 component in the intensity map, a kinematic connection between W-4 and the bridge cannot be explained in our simple kinematic model. Nevertheless, we cannot rule out the possibility that the W-4 component and the western streamer are formed from the 20 MC since the kinematic structures might have been washed out during the passage of the 20 MC in the central few pc \citep{takekawa17}.

We also find that the western streamers intersect with the CND in Figure~\ref{fig-phase-cnd1} and ~\ref{fig-phase-cnd2}, which provides  some evidence that the western streamers are ``converging and feeding'' toward the CND. For the eastern streamer, however, we notice that it is fainter than the western streamers and it does not show a clear signature of transition into the CND. One of the possible reasons is that the feedback of the supernova remnant (SNR) SgrA East \citep{mezger,ekers} may disturb the eastern streamer. The SgrA East is known to interact with the 50 MC \citep[e.g.,][]{genzel90,ho91,serabyn92,zylka99}. The age of the SgrA East ($10^{4}$ years) \citep{mcgary01} is consistent with the dynamical time scale of the eastern streamer.
Owing to the large velocity gradient and elevated temperature of the western streamer, \citet{mcgary01} also proposed that the western streamer appears to have been swept up by the expanding shell of the SgrA East. The western streamer shows expanding motion outwards with the SNR shell.
In this regard, the ''infall velocity'' implemented in our simple model may also be interpreted as an expanding motion. However, we do not see significant morphological and kinematic disturbances of the western streamer in our SMA and NRO 45-m data.

In this paper, we alternatively propose that the radial motion of the eastern/western streamers may be attributed to the infall toward the CND.
The infall picture based on our data is sketched in Figure~\ref{fig-model}. The molecular gas originated from the ambient clouds is tidally stretched into long streamers (protrusions). These streamers are gravitationally captured by the central potential and spirally rotate as they converge toward the CND. This picture is consistent with the past theoretical modelings, which predict that the CND is formed by the tidal capture and disruption of nearby molecular clouds \citep{sanders98,vollmer02,wardle08,mapelli16}.

\subsection{Lower limit of the infall rate}

We estimate the mass inflow rate based on the NRO 45-m data because the SMA map suffers from missing flux. We estimate the infall rate of the eastern streamer assuming that it is currently stripped away from the 50 MC and approaching the CND.
The molecular gas mass of the eastern streamer under local thermal equilibrium (LTE) assumption is $\sim120~M_{\odot}$ (CS($J=2-1$)/($J=1-0$) intensity ratio of 1). The  mass accretion rate over $7\times10^{4}$ years is thus $\sim2\times10^{-3}~M_{\odot}~\rm yr^{-1}$. This accretion rate is consistent with the simulation in \citet{vollmer02}. The total accreting gas mass is therefore $\sim200~{\rm M}_{\odot}$ over $10^{5}$ year
if the accretion rate is consistent. The life time of the CND is a long-standing issue \citep{guesten87,chris,maria,great,mills13b,smith14,harada15}. The CND may be transient and will dissolve if its density is lower than the tidal threshold ($\sim10^{7}$ cm$^{-3}$). In this paper, we find that the replenishment of the gas accreting onto the CND is on going, we then estimate whether this accretion rate is significant to sustain the life time of the CND.
The gas mass of the CND determined by molecular gas and dust has a larger discrepancy from 10$^{4}~M_{\odot}$ to 10$^{6}~M_{\odot}$ \citep{chris,maria,genzel10,great,etx}. The mass ratio of accreting to CND mass is then around 0.02-2\%.
This number is a lower limit since there are other substructures in the 50 MC showing infalling signatures in the lv-diagrams. Moreover, we notice that the seeding material in the 50 MC is more from the compact clouds rather than from the diffuse envelopes of the 50 MC. The individual clump masses in the 50 MC are in the range of $100-1000~M_{\odot}$ \citep{tsuboi12}. The clump densities are on the order of $10^{3-4}$ cm$^{-3}$, which is lower than the tidal threshold limit. These observational results suggest that the 50 MC is a possible ``seeding factory'' to continue feeding the CND over its crossing time (a few $10^{5}$ years to cross 20 pc).
We therefore conclude that in spite of the density arguments on the life time of the CND \citep{chris,maria,great,mills13b,smith14,harada15}, a precise measurement of the CND mass will be also crucial to determine its life time.

\section{Conclusions}

In this paper, we present a dynamical picture of the infall scenario connecting from 20 pc to the central 2 pc, which might also have implications for the nuclei feeding in nearby galaxies.
With our wide-field map of CS($J=2-1$), we find that the western streamer identified in the interferometric maps has a counterpart called extW-streamer. We also find a feature called eastern streamer in CS($J=2-1$) which shows a physical connection to the 50 MC via tidally stretched clouds. The extW- and eastern streamers are the outskirt of the CND and show a slower rotation than the CND. 
We also find that the extW- and the eastern streamers can be described by a simple Keplerian rotation and infall model. The extW-streamer (western streamer) also shows kinematic and morphological evidence of an intersection with the CND, supplying replenishing gas for the CND. If the accretion rate is constant over $10^{5}$ years (the crossing time of the 50 MC), then the lower limits of the accretion-to-mass fractions onto the CND are in the range of 0.02-2\%. An accurate measurement of the CND mass will be essential to evaluate the life time of the CND.

\acknowledgements

We thank the reviewer for a thoughtful review and constructive comments to improve the manuscript.
Pei-Ying Hsieh is supported by the Ministry of Science and Technology (MoST) of Taiwan through the grants MOST 105-2112-M-001-025-MY3, MOST 104-2119-M-001-019-MY3, MOST 105-2119-M-001-042, and MOST 106-2119-M-001-013. Patrick M. K. acknowledges support from MoST 103-2119-M-001-009 and from an Academia Sinica Career Award. W.-T.K. was supported by the National Research Foundation of Korea (NRF) grant funded by the Korea government (MEST) (No.~3348-20160021). Chorng-Yuan Hwang acknowledges support from MoST 106-2119-M-008-016.

%\nocite{*}
%\bibliography{inflow.ref}

\clearpage

\begin{thebibliography}{}
\expandafter\ifx\csname natexlab\endcsname\relax\def\natexlab#1{#1}\fi
\providecommand{\url}[1]{\href{#1}{#1}}

\bibitem[{{Amo-Baladr{\'o}n} {et~al.}(2011){Amo-Baladr{\'o}n},
  {Mart{\'{\i}}n-Pintado}, \& {Mart{\'{\i}}n}}]{amo11}
{Amo-Baladr{\'o}n}, M.~A., {Mart{\'{\i}}n-Pintado}, J., \& {Mart{\'{\i}}n}, S.
  2011, \aap, 526, A54

\bibitem[{{Christopher} {et~al.}(2005){Christopher}, {Scoville}, {Stolovy}, \&
  {Yun}}]{chris}
{Christopher}, M.~H., {Scoville}, N.~Z., {Stolovy}, S.~R., \& {Yun}, M.~S.
  2005, \apj, 622, 346

\bibitem[{{Coil} \& {Ho}(2000)}]{coil00}
{Coil}, A.~L., \& {Ho}, P.~T.~P. 2000, \apj, 533, 245

\bibitem[{{Ekers} {et~al.}(1983){Ekers}, {van Gorkom}, {Schwarz}, \&
  {Goss}}]{ekers}
{Ekers}, R.~D., {van Gorkom}, J.~H., {Schwarz}, U.~J., \& {Goss}, W.~M. 1983,
  \aap, 122, 143

\bibitem[{{Etxaluze} {et~al.}(2011){Etxaluze}, {Smith}, {Tolls}, {Stark}, \&
  {Gonz{\'a}lez-Alfonso}}]{etx}
{Etxaluze}, M., {Smith}, H.~A., {Tolls}, V., {Stark}, A.~A., \&
  {Gonz{\'a}lez-Alfonso}, E. 2011, \aj, 142, 134

\bibitem[{{Genzel} {et~al.}(2010){Genzel}, {Eisenhauer}, \&
  {Gillessen}}]{genzel10}
{Genzel}, R., {Eisenhauer}, F., \& {Gillessen}, S. 2010, Reviews of Modern
  Physics, 82, 3121

\bibitem[{{Genzel} {et~al.}(1990){Genzel}, {Stacey}, {Harris}, {Townes},
  {Geis}, {Graf}, {Poglitsch}, \& {Stutzki}}]{genzel90}
{Genzel}, R., {Stacey}, G.~J., {Harris}, A.~I., {et~al.} 1990, \apj, 356, 160

\bibitem[{{Ghez} {et~al.}(2005){Ghez}, {Salim}, {Hornstein}, {Tanner}, {Lu},
  {Morris}, {Becklin}, \& {Duch{\^e}ne}}]{ghez05}
{Ghez}, A.~M., {Salim}, S., {Hornstein}, S.~D., {et~al.} 2005, \apj, 620, 744

\bibitem[{{Guesten} {et~al.}(1987){Guesten}, {Genzel}, {Wright}, {Jaffe},
  {Stutzki}, \& {Harris}}]{guesten87}
{Guesten}, R., {Genzel}, R., {Wright}, M.~C.~H., {et~al.} 1987, \apj, 318, 124

\bibitem[{{Harada} {et~al.}(2015){Harada}, {Riquelme}, {Viti},
  {Jim{\'e}nez-Serra}, {Requena-Torres}, {Menten}, {Mart{\'{\i}}n}, {Aladro},
  {Martin-Pintado}, \& {Hochg{\"u}rtel}}]{harada15}
{Harada}, N., {Riquelme}, D., {Viti}, S., {et~al.} 2015, \aap, 584, A102

\bibitem[{{Harris} {et~al.}(1985){Harris}, {Jaffe}, {Silber}, \&
  {Genzel}}]{harris}
{Harris}, A.~I., {Jaffe}, D.~T., {Silber}, M., \& {Genzel}, R. 1985, \apjl,
  294, L93

\bibitem[{{Herrnstein} \& {Ho}(2002)}]{herrnstein02}
{Herrnstein}, R.~M., \& {Ho}, P.~T.~P. 2002, \apjl, 579, L83

\bibitem[{{Herrnstein} \& {Ho}(2005{\natexlab{a}})}]{herrnstein05}
---. 2005{\natexlab{a}}, \apj, 620, 287

\bibitem[{{Herrnstein} \& {Ho}(2005{\natexlab{b}})}]{nh3_05}
---. 2005{\natexlab{b}}, \apj, 620, 287

\bibitem[{{Ho} {et~al.}(1991){Ho}, {Ho}, {Szczepanski}, {Jackson}, \&
  {Armstrong}}]{ho91}
{Ho}, P.~T.~P., {Ho}, L.~C., {Szczepanski}, J.~C., {Jackson}, J.~M., \&
  {Armstrong}, J.~T. 1991, \nat, 350, 309

\bibitem[{{Hsieh} {et~al.}(2015){Hsieh}, {Ho}, \& {Hwang}}]{hsieh15}
{Hsieh}, P.-Y., {Ho}, P.~T.~P., \& {Hwang}, C.-Y. 2015, \apj, 811, 142

\bibitem[{{Hsieh} {et~al.}(2016){Hsieh}, {Ho}, {Hwang}, {Shimajiri},
  {Matsushita}, {Koch}, \& {Iono}}]{hsieh16}
{Hsieh}, P.-Y., {Ho}, P.~T.~P., {Hwang}, C.-Y., {et~al.} 2016, \apj, 831, 72

\bibitem[{{Jackson} {et~al.}(1993){Jackson}, {Geis}, {Genzel}, {Harris},
  {Madden}, {Poglitsch}, {Stacey}, \& {Townes}}]{jackson93}
{Jackson}, J.~M., {Geis}, N., {Genzel}, R., {et~al.} 1993, \apj, 402, 173

\bibitem[{{Lau} {et~al.}(2013){Lau}, {Herter}, {Morris}, {Becklin}, \&
  {Adams}}]{lau}
{Lau}, R.~M., {Herter}, T.~L., {Morris}, M.~R., {Becklin}, E.~E., \& {Adams},
  J.~D. 2013, \apj, 775, 37

\bibitem[{{Lee} {et~al.}(2003){Lee}, {Pak}, {Davis}, {Herrnstein}, {Geballe},
  {Ho}, \& {Wheeler}}]{lee03}
{Lee}, S., {Pak}, S., {Davis}, C.~J., {et~al.} 2003, \mnras, 341, 509

\bibitem[{{Liu} {et~al.}(2012){Liu}, {Hsieh}, {Ho}, {Su}, {Wright}, {Sun}, \&
  {Minh}}]{liu12}
{Liu}, H.~B., {Hsieh}, P.-Y., {Ho}, P.~T.~P., {et~al.} 2012, \apj, 756, 195

\bibitem[{{Mapelli} \& {Trani}(2016)}]{mapelli16}
{Mapelli}, M., \& {Trani}, A.~A. 2016, \aap, 585, A161


\bibitem[{{Mart{\'{\i}}n} {et~al.}(2012){Mart{\'{\i}}n},
  {Mart{\'{\i}}n-Pintado}, {Montero-Casta{\~n}o}, {Ho}, \& {Blundell}}]{martin12}
---. 2012{\natexlab{b}}, \aap, 539, A29

\bibitem[{{McGary} {et~al.}(2001){McGary}, {Coil}, \& {Ho}}]{mcgary01}
{McGary}, R.~S., {Coil}, A.~L., \& {Ho}, P.~T.~P. 2001, \apj, 559, 326

\bibitem[{{Mezger} {et~al.}(1989){Mezger}, {Zylka}, {Salter}, {Wink}, {Chini},
  {Kreysa}, \& {Tuffs}}]{mezger}
{Mezger}, P.~G., {Zylka}, R., {Salter}, C.~J., {et~al.} 1989, \aap, 209, 337

\bibitem[{{Mills} {et~al.}(2013){Mills}, {G{\"u}sten}, {Requena-Torres}, \&
  {Morris}}]{mills13b}
{Mills}, E.~A.~C., {G{\"u}sten}, R., {Requena-Torres}, M.~A., \& {Morris},
  M.~R. 2013, \apj, 779, 47

\bibitem[{{Montero-Casta{\~n}o} {et~al.}(2009){Montero-Casta{\~n}o},
  {Herrnstein}, \& {Ho}}]{maria}
{Montero-Casta{\~n}o}, M., {Herrnstein}, R.~M., \& {Ho}, P.~T.~P. 2009, \apj,
  695, 1477

\bibitem[{{Oka} {et~al.}(2011){Oka}, {Nagai}, {Kamegai}, \& {Tanaka}}]{oka11}
{Oka}, T., {Nagai}, M., {Kamegai}, K., \& {Tanaka}, K. 2011, \apj, 732, 120

\bibitem[{{Okumura} {et~al.}(1989){Okumura}, {Ishiguro}, {Fomalont}, {Chikada},
  {Kasuga}, {Morita}, {Kawabe}, {Kobayashi}, {Kanzawa}, {Iwashita}, \&
  {Hasegawa}}]{okumura89}
{Okumura}, S.~K., {Ishiguro}, M., {Fomalont}, E.~B., {et~al.} 1989, \apj, 347,
  240

\bibitem[{{Requena-Torres} {et~al.}(2012){Requena-Torres}, {G{\"u}sten},
  {Wei{\ss}}, {Harris}, {Mart{\'{\i}}n-Pintado}, {Stutzki}, {Klein},
  {Heyminck}, \& {Risacher}}]{great}
{Requena-Torres}, M.~A., {G{\"u}sten}, R., {Wei{\ss}}, A., {et~al.} 2012, \aap,
  542, L21

\bibitem[{{Roberts} \& {Goss}(1993)}]{roberts93}
{Roberts}, D.~A., \& {Goss}, W.~M. 1993, \apjs, 86, 133

\bibitem[{{Sanders}(1998)}]{sanders98}
{Sanders}, R.~H. 1998, \mnras, 294, 35

\bibitem[{{Serabyn} {et~al.}(1992){Serabyn}, {Lacy}, \&
  {Achtermann}}]{serabyn92}
{Serabyn}, E., {Lacy}, J.~H., \& {Achtermann}, J.~M. 1992, \apj, 395, 166

\bibitem[{{Smith} \& {Wardle}(2014)}]{smith14}
{Smith}, I.~L., \& {Wardle}, M. 2014, \mnras, 437, 3159

\bibitem[{{Takekawa} {et~al.}(2017){Takekawa}, {Oka}, \& {Tanaka}}]{takekawa17}
{Takekawa}, S., {Oka}, T., \& {Tanaka}, K. 2017, \apj, 834, 121

\bibitem[{{Tsuboi} {et~al.}(1999){Tsuboi}, {Handa}, \& {Ukita}}]{tsuboi99}
{Tsuboi}, M., {Handa}, T., \& {Ukita}, N. 1999, \apjs, 120, 1

\bibitem[{{Tsuboi} \& {Miyazaki}(2012)}]{tsuboi12}
{Tsuboi}, M., \& {Miyazaki}, A. 2012, \pasj, 64, 111

\bibitem[{{Vollmer} \& {Duschl}(2002)}]{vollmer02}
{Vollmer}, B., \& {Duschl}, W.~J. 2002, \aap, 388, 128

\bibitem[{{Wardle} \& {Yusef-Zadeh}(2008)}]{wardle08}
{Wardle}, M., \& {Yusef-Zadeh}, F. 2008, \apjl, 683, L37

\bibitem[{{Wright} {et~al.}(2001){Wright}, {Coil}, {McGary}, {Ho}, \&
  {Harris}}]{wright}
{Wright}, M.~C.~H., {Coil}, A.~L., {McGary}, R.~S., {Ho}, P.~T.~P., \&
  {Harris}, A.~I. 2001, \apj, 551, 254

\bibitem[{{Zylka} {et~al.}(1999){Zylka}, {G{\"u}sten}, {Philipp}, {Ungerechts},
  {Mezger}, \& {Duschl}}]{zylka99}
{Zylka}, R., {G{\"u}sten}, R., {Philipp}, S., {et~al.} 1999, in Astronomical
  Society of the Pacific Conference Series, Vol. 186, The Central Parsecs of
  the Galaxy, ed. H.~{Falcke}, A.~{Cotera}, W.~J. {Duschl}, F.~{Melia}, \&
  M.~J. {Rieke}, 415

\end{thebibliography}

\begin{figure}
\begin{center}
\epsscale{1}
\includegraphics[angle=0,scale=0.7]{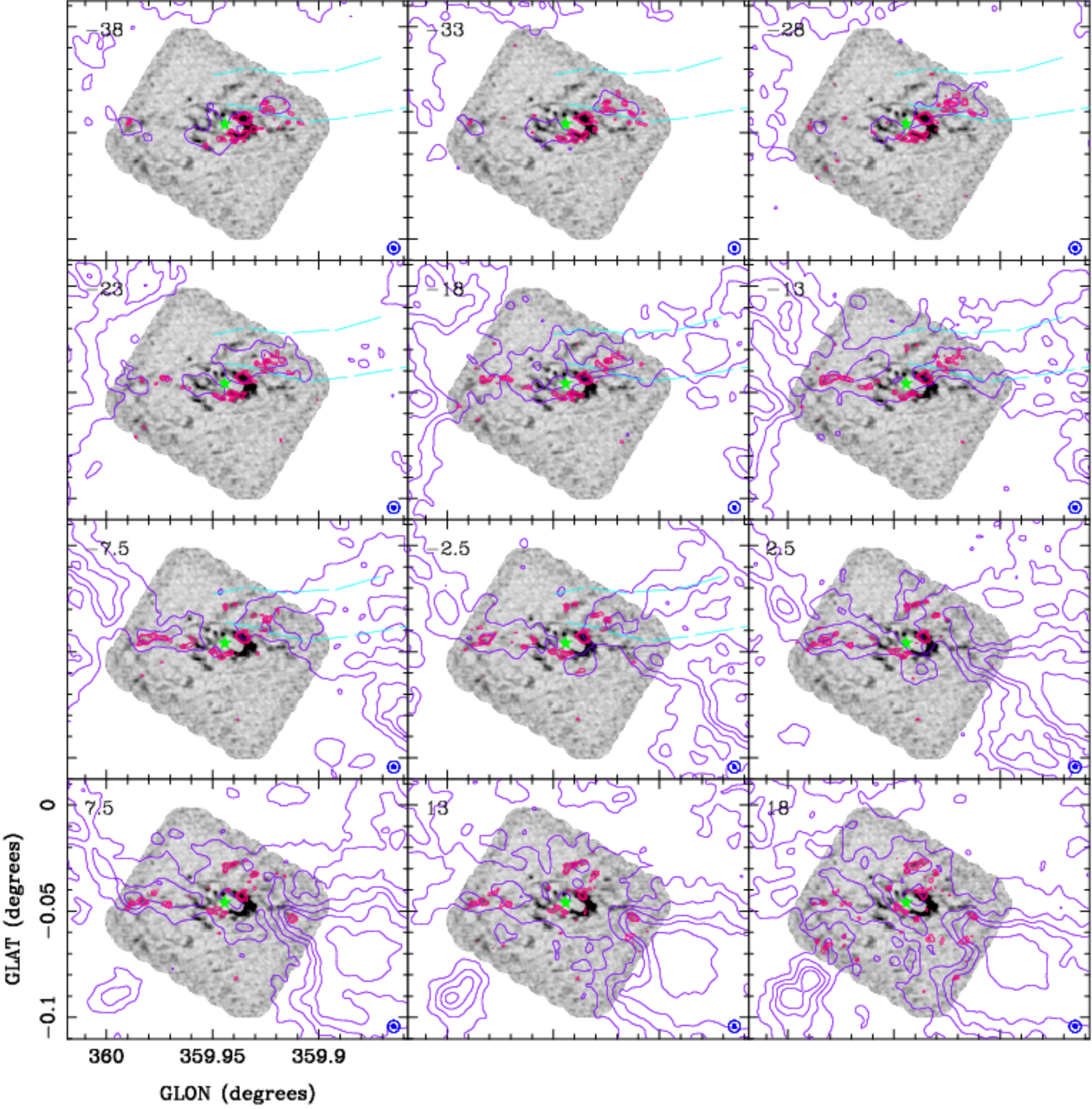}
\caption[]{Channel maps of our CS($J=2-1$) line (purple contours) and the SMA HCN($J=4-3$) line (magenta contours) overlaid on the integrated intensity HCN($J=4-3$) map (grey scale). The contour levels of CS are, 5, 10, 15, 20, 30$\times$0.06 K, and they are 3, 5, 10$\times$1 Jy beam$^{-1}$ for HCN.
The radial velocity with respect to the systematic velocity is labeled on the top left corner. The blue circle and filled ellipse show the beam sizes of the CS (20$\arcsec$) and the HCN ($5.87\arcsec \times 4.44\arcsec$) maps. The extW-streamer and the western streamer are located in the region bounded by two cyan lines. SgrA* is labeled with the green star.}
\label{fig-chan}
\end{center}
\end{figure}

\begin{figure}
\begin{center}
\epsscale{1}
\includegraphics[angle=0,scale=0.7]{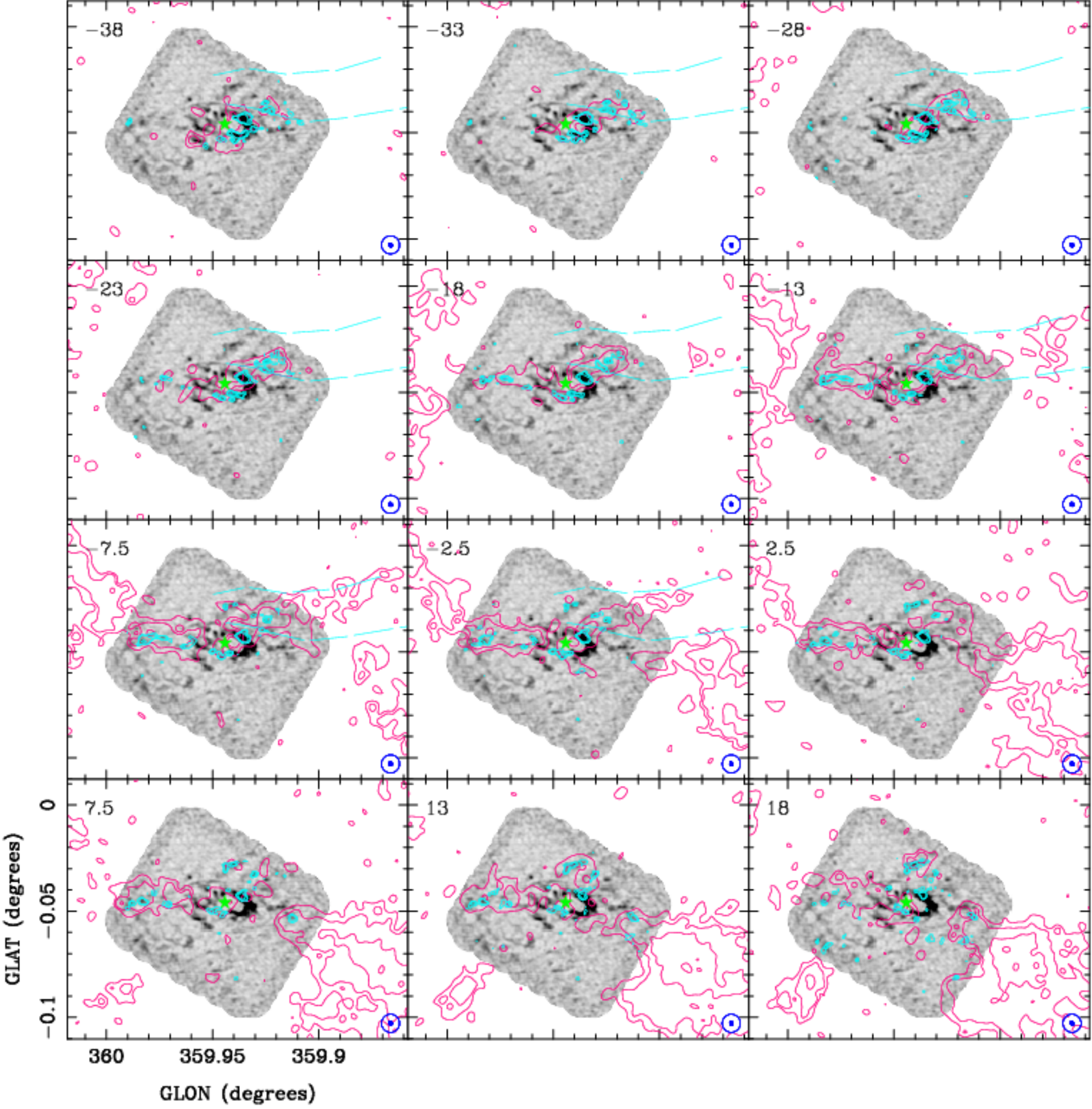}
\caption[]{Channel maps of our CS($J=5-4$) line (rose contours) and the SMA HCN($J=4-3$) line (cyan contours) overlaid on the integrated intensity HCN($J=4-3$) map (grey scale). The contour levels of CS are, 3, 5, 10, 15, 20, 30$\times$0.12 K, and they are 3, 5, 10$\times$1 Jy beam$^{-1}$ for HCN. The beam size of CS($J=5-4$) is 31$\arcsec$.
The radial velocity with respect to the systematic velocity is labeled on the top left corner. The blue circle and filled ellipse show the beam sizes of the CS (31$\arcsec$) and the HCN ($5.87\arcsec \times 4.44\arcsec$) maps. The extW-streamer and the western streamer are located in the region bounded by two cyan lines. SgrA* is labeled with the green star.}
\label{fig-chan}
\end{center}
\end{figure}

\begin{figure}
\begin{center}
\epsscale{0.5}
\includegraphics[scale=0.4]{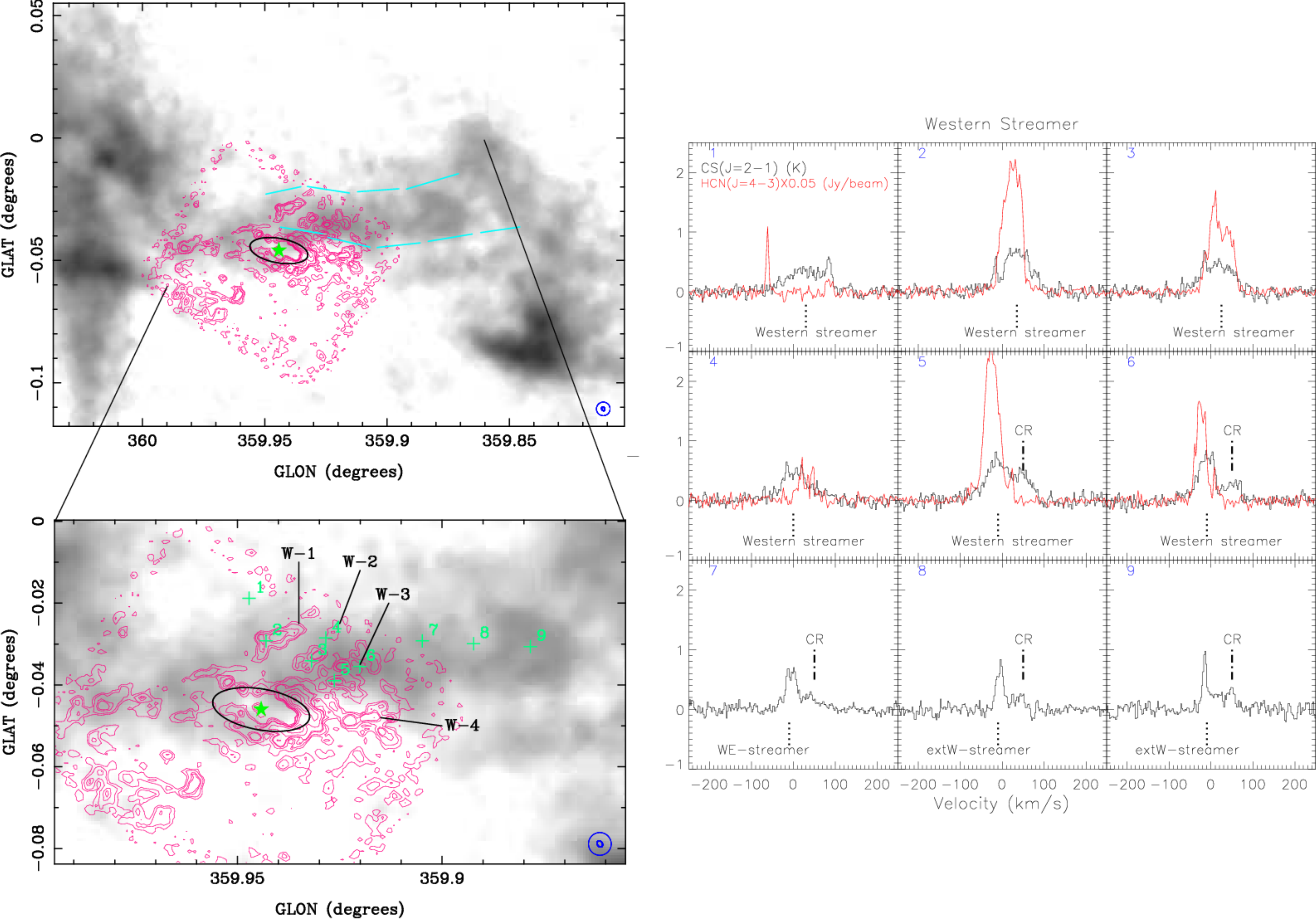}
\caption[]{CS($J=2-1$) line integrated from $-24$ km s$^{-1}$ to $-11$ km s$^{-1}$ (grey) overlaid on the HCN($J=4-3$) map (integrated from $-160$ km s$^{-1}$ to 150 km s$^{-1}$ in magenta contours). The contours are  2, 4, 8, 10, 20, 40, 60, 80, 100$\times$10 Jy beam$^{-1}$ km s$^{-1}$. SgrA* is labeled with the green star. The extW-streamer and the western streamer are located in the region bounded by two cyan lines. The locations of the W-1 to W-4 and the CND (black ellipse) are marked. The spectra of the corresponding positions are marked. Red and black lines are the spectral data of the HCN($J=4-3$) and CS($J=2-1$), respectively. The spectral components of the extW-streamer, western streamer, and the connecting ridge (CR) \citep{hsieh15} are labeled.
}
\label{fig-west}
\end{center}
\end{figure}

\begin{figure}
\epsscale{0.2}
\begin{center}
\includegraphics[angle=0,scale=0.4]{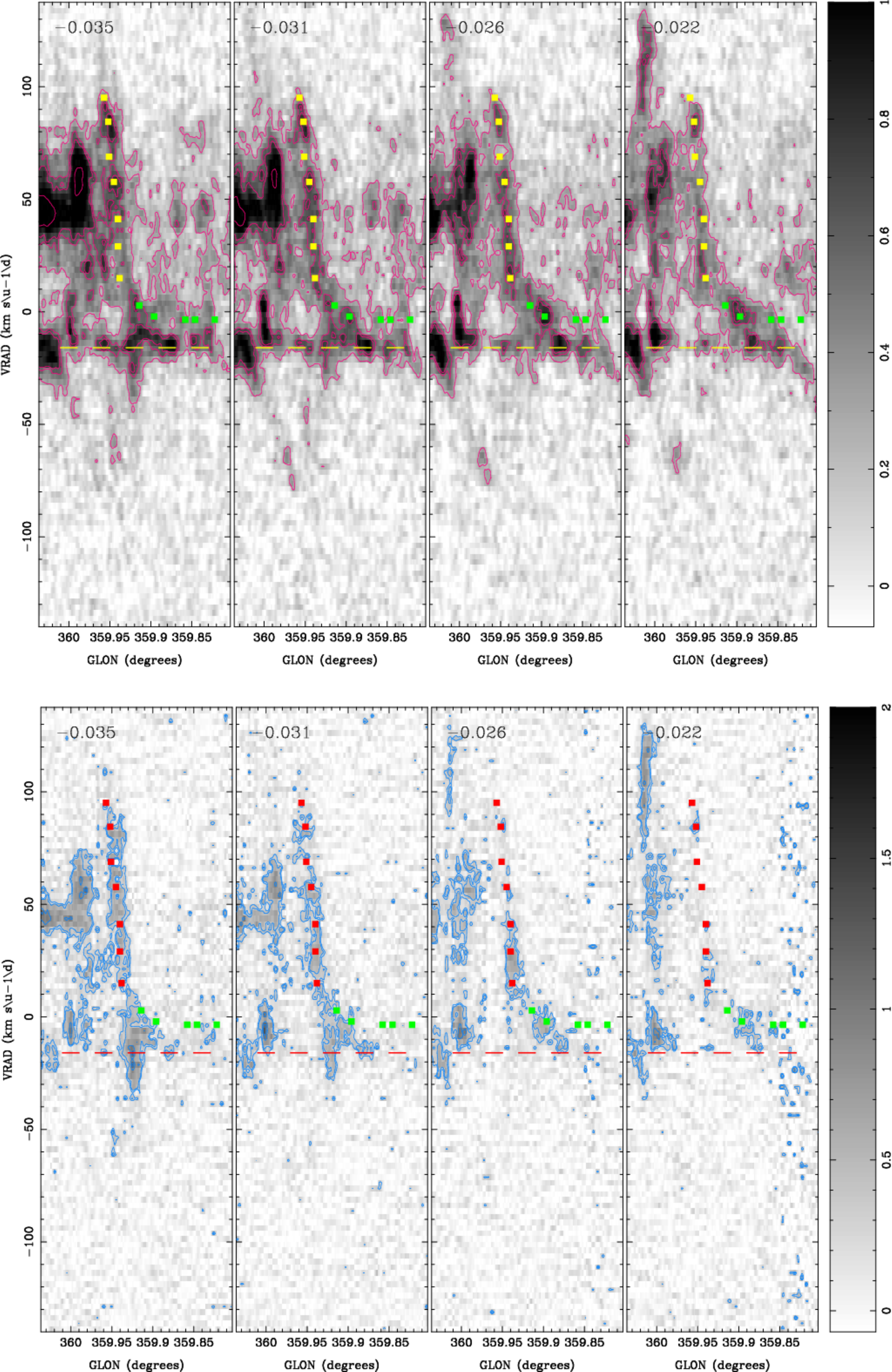}
\caption[]{Upper: CS($J=2-1$) $lv$-diagrams of the ext-W and western streamers. The contours are 4, 8, 10, 20$\times0.07$ K ($T_{\rm A}^*$). The position of the latitude is labeled on the top left corner. The green squares mark our definition of the extW-streamer and the yellow squares mark the western streamer. The ``$-16$'' km s$^{-1}$ component is labeled with the yellow lines. Lower: CS($J=5-4$) $lv$-diagrams of the ext-W and western streamers.  The contours are 3, 4, 8$\times$0.11 K ($T_{\rm A}^*$). The $-16$ km s$^{-1}$ component is labeled with the red lines. The red squares mark the western streamer.
}
\label{fig-pv-west}
\end{center}
\end{figure}

\begin{figure}
\epsscale{0.2}
\begin{center}
\includegraphics[angle=0,scale=0.4]{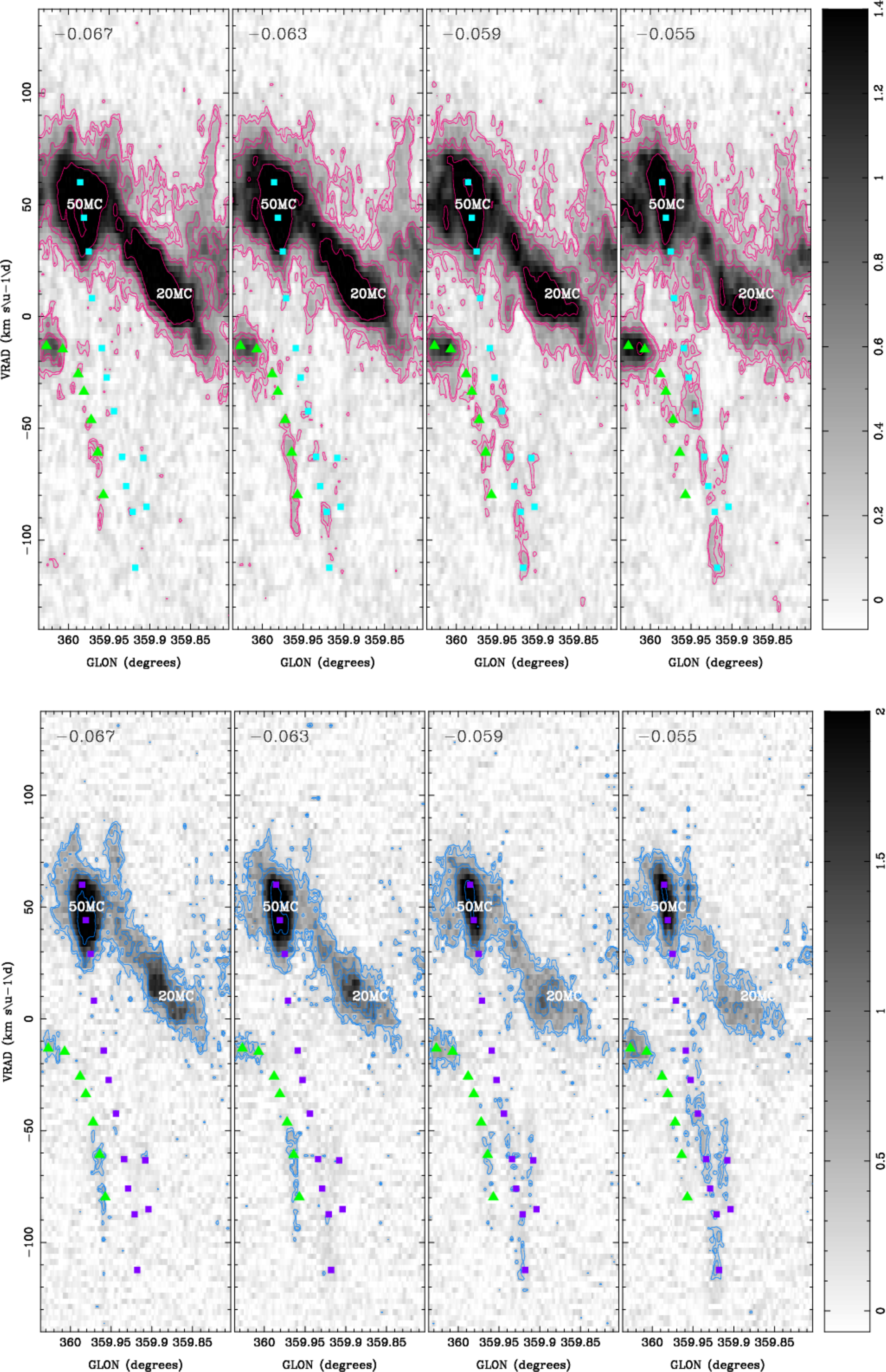}
% average from b=ch 73 to ch=91
\caption[]{Upper: CS($J=2-1$) $lv$-diagrams of the eastern streamer.
The contours are 3, 4, 8, 10, 20$\times0.07$ K ($T_{\rm A}^*$).  The position of the latitude is labeled on the top left corner. The cyan squares mark the eastern streamer and the green triangles mark the feature ``C1'' mentioned in \citet{oka11}. Lower: CS($J=5-4$) $lv$-diagrams of the eastern streamers.  The contours are 3, 4, 8$\times$0.11 K ($T_{\rm A}^*$). The purple squares mark the eastern streamer.
}
\label{fig-pv-east}
\end{center}
\end{figure}

\begin{figure}
\begin{center}
\epsscale{0.4}
\includegraphics[angle=0,scale=0.4]{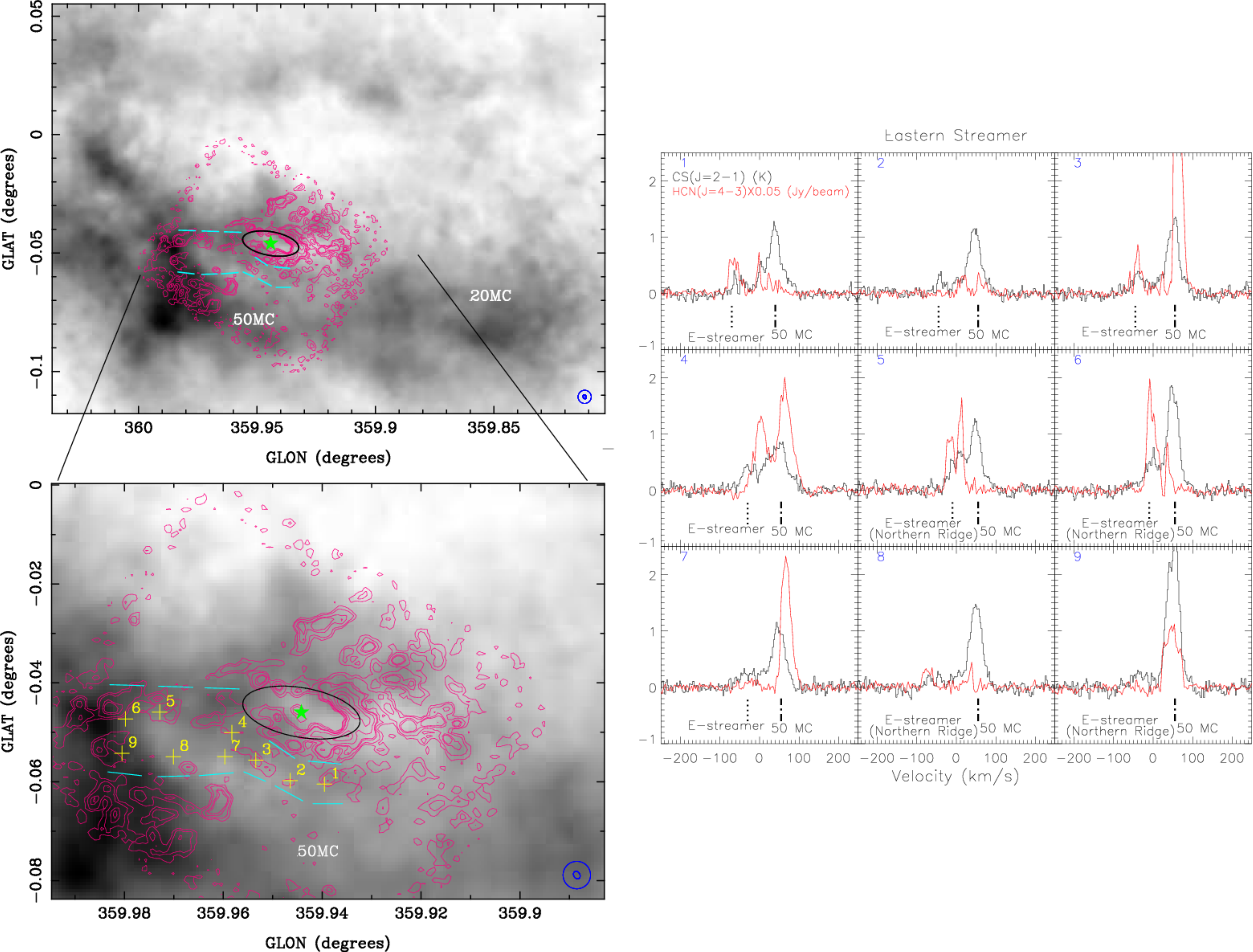}
\caption[]{CS($J=2-1$) line integrated from $-175$ km s$^{-1}$ to $175$ km s$^{-1}$ (grey) overlaid on the HCN($J=4-3$) map (magenta contours). The contours are, 2, 4, 8, 10, 20, 40, 60, 80, 100$\times$10 Jy beam$^{-1}$ km s$^{-1}$. SgrA* is labeled with the green star. The eastern streamer is in the regions bounded by the cyan lines. The spectra of the corresponding positions are marked.   Red and black lines are spectral data of HCN($J=4-3$) and CS($J=2-1$), respectively. The spectral components of the eastern streamer, the northern ridge, and the 50 MC are labeled.
}
\label{fig-east}
\end{center}
\end{figure}

\begin{figure}
\begin{center}
\includegraphics[scale=0.3]{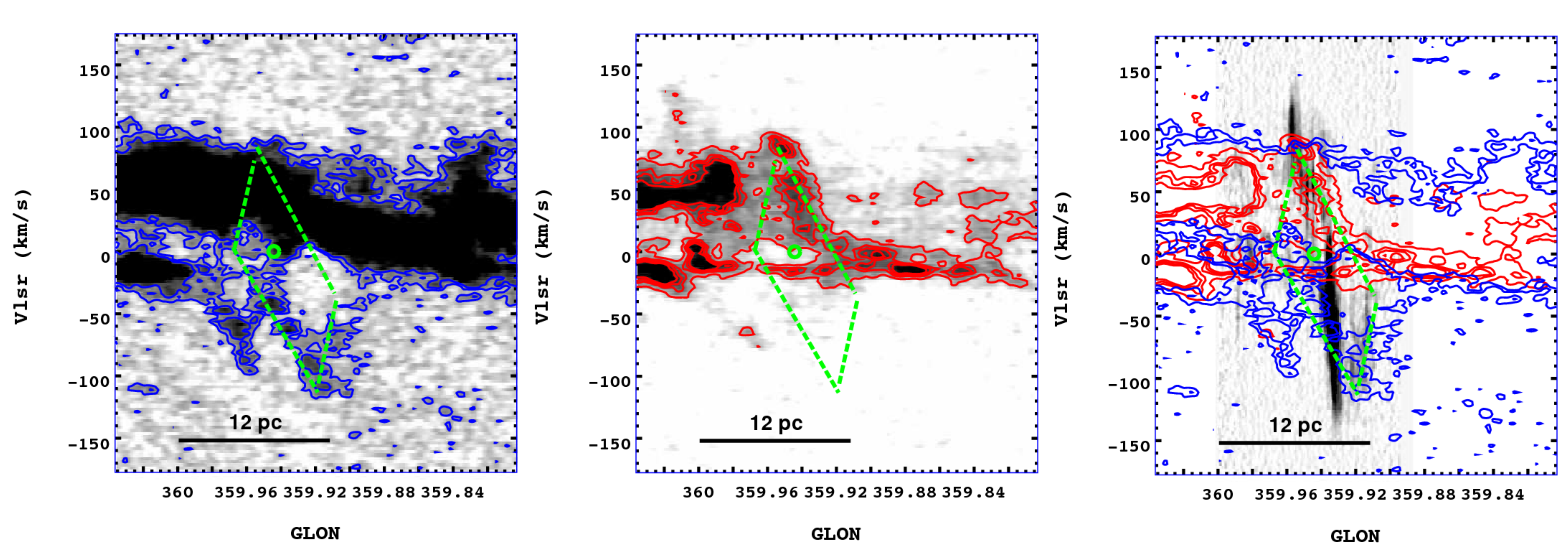}
\caption[]{Left: $lv$-diagrams of CS($J=2-1$) eastern streamer; middle: extW/western streamer; right: extW/western streamer (red contours) and eastern streamer (blue contours) overlaid on the HCN($J=4-3$) (CND: grey). The location of SgrA* is labeled as small circle. The green segments label the ``parallelogram'' tracing non-circular motions of the streamers (also see \citet{oka11}).
}
\label{fig-pv-two-ring}
\end{center}
\end{figure}

\begin{figure}
\begin{center}
%\epsscale{0.6}
\includegraphics[angle=0,scale=0.4]{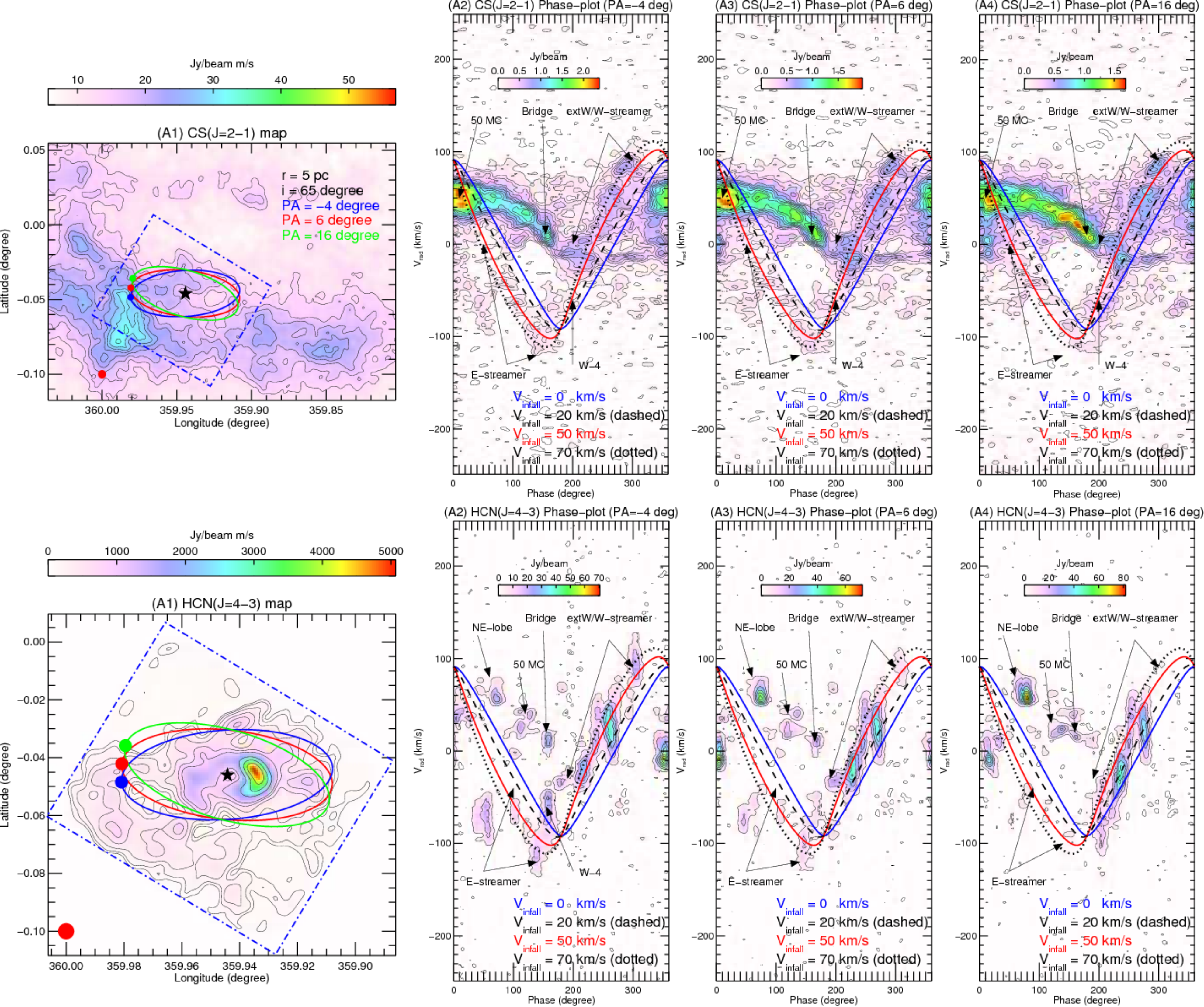}
\caption[]{Phase plots of the streamers.
 Upper: (A1) CS($J=2-1$) integrated intensity map. The beam size of 20$\arcsec$ is shown with the red dot in the lower left corner. The blue, red, and green elliptical annuli mark the trajectory of the phase plot (A2, A3, A4) with position angle of $-4\degr$, 6$\degr$, and 16$\degr$, respectively. The intrinsic radius is 5 pc. Starting position ($0\degr$) of the trajectory is labeled by the dots, and the phase is counterclockwise. The blue box marks the map size of the SMA data. Pure Keplerian ($V_{\rm infall}=0$ km s$^{-1}$; blue curve) and Keplerian plus infall models are overlaid on (A2-A4) with a radius of 5 pc, an inclination angle of 65$\degr$, and the PA of $-4\degr$, 6$\degr$, and 16$\degr$. The infall velocities are 20 km s$^{-1}$ (dashed line), 50 km s$^{-1}$ (red line) and 70 km s$^{-1}$ (dotted line), respectively. Lower: Same as above but for the HCN($J=4-3$) data.
}
\label{fig-phase-st1}
\end{center}
\end{figure}

\begin{figure}
\begin{center}
\includegraphics[angle=0,scale=0.4]{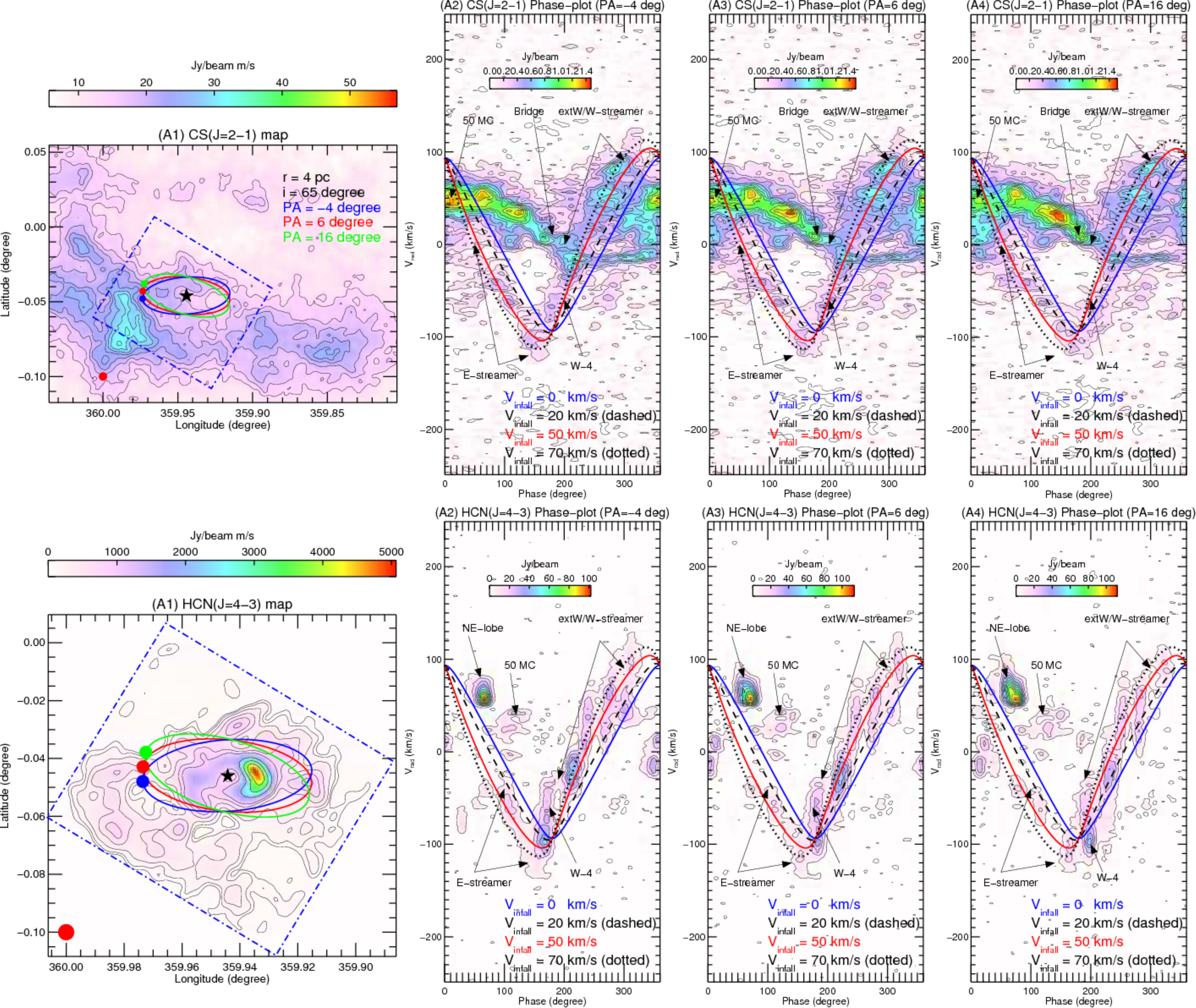}
\caption[]{Phase plots of the streamers.
 Upper: (A1) CS($J=2-1$) integrated intensity map. The beam size of 20$\arcsec$ is shown with the red dot in the lower left corner. The blue, red, and green elliptical annuli mark the trajectory of the phase plot (A2, A3, A4) with position angle of $-4\degr$, 6$\degr$, and 16$\degr$, respectively. The intrinsic radius is 4 pc. Starting position ($0\degr$) of the trajectory is labeled by the dots, and the phase is counterclockwise. The blue box marks the map size of the SMA data. Pure Keplerian ($V_{\rm infall}=0$ km s$^{-1}$; blue curve) and Keplerian plus infall models are overlaid on (A2-A4) with a radius of 4 pc, an inclination angle of 65$\degr$, and the PA of $-4\degr$, 6$\degr$, and 16$\degr$. The infall velocities are 20 km s$^{-1}$ (dashed line), 50 km s$^{-1}$ (red line) and 70 km s$^{-1}$ (dotted line), respectively. Lower: Same as above but for the HCN($J=4-3$) data.
}
\label{fig-phase-st2}
\end{center}
\end{figure}

\begin{figure}
\begin{center}
\includegraphics[angle=0,scale=0.4]{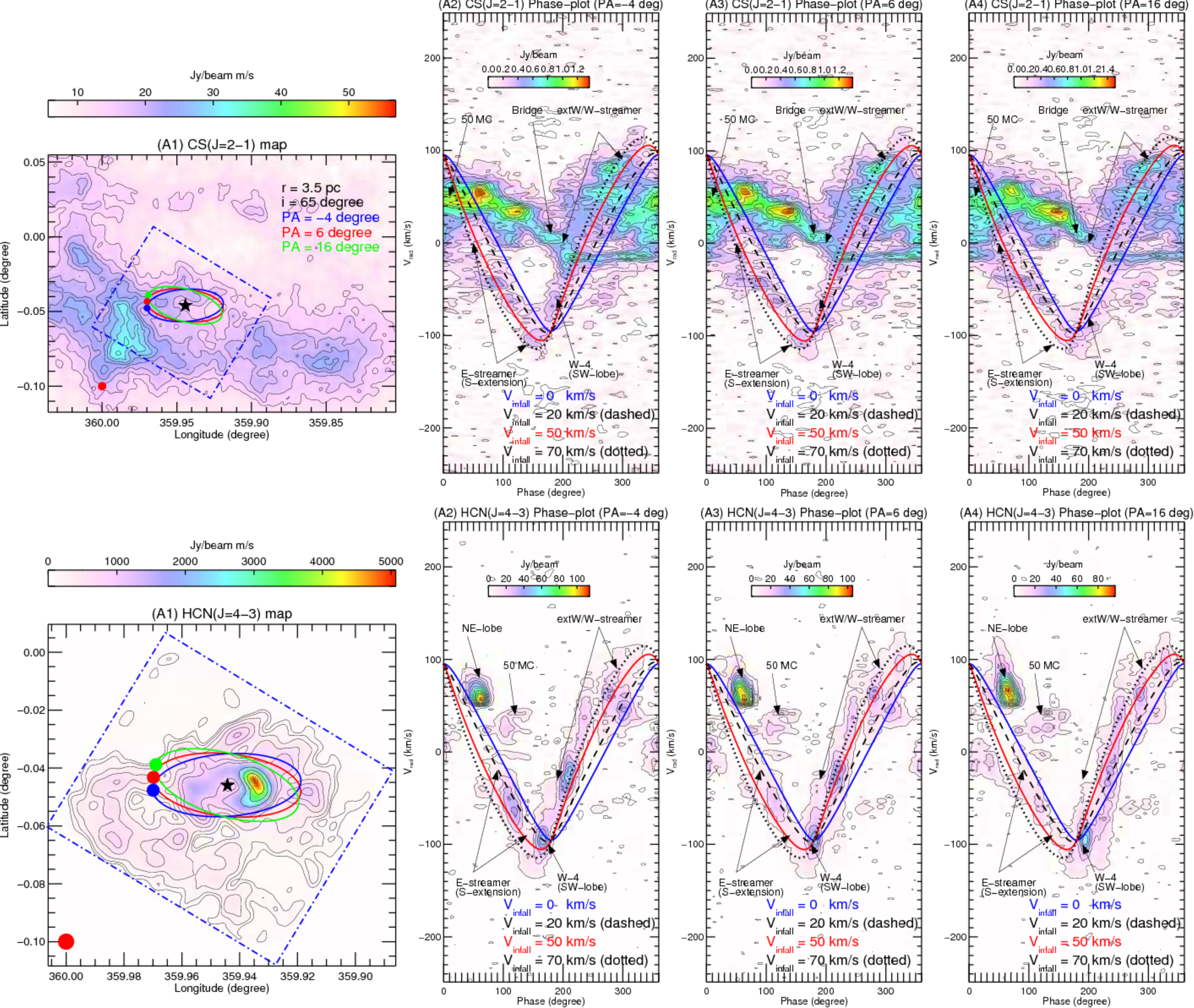}
\caption[]{Phase plots of the CND.
 Upper: (A1) CS($J=2-1$) integrated intensity map. The beam size of 20$\arcsec$ is shown with the red dot in the lower left corner. The blue, red, and green elliptical annuli mark the trajectory of the phase plot (A2, A3, A4) with position angle of $-4\degr$, 6$\degr$, and 16$\degr$, respectively. The intrinsic radius is 3.5 pc. Starting position ($0\degr$) of the trajectory is labeled by the dots, and the phase is counterclockwise. The blue box marks the map size of the SMA data. Pure Keplerian ($V_{\rm infall}=0$ km s$^{-1}$; blue curve) and Keplerian plus infall models are overlaid on (A2-A4) with a radius of 3.5 pc, an inclination angle of 65$\degr$, and the PA of $-4\degr$, 6$\degr$, and 16$\degr$. The infall velocities are 20 km s$^{-1}$ (dashed line), 50 km s$^{-1}$ (red line) and 70 km s$^{-1}$ (dotted line), respectively. Lower: Same as above but for the HCN($J=4-3$) data.
}
\label{fig-phase-cnd1}
\end{center}
\end{figure}

\begin{figure}
\begin{center}
\includegraphics[angle=0,scale=0.4]{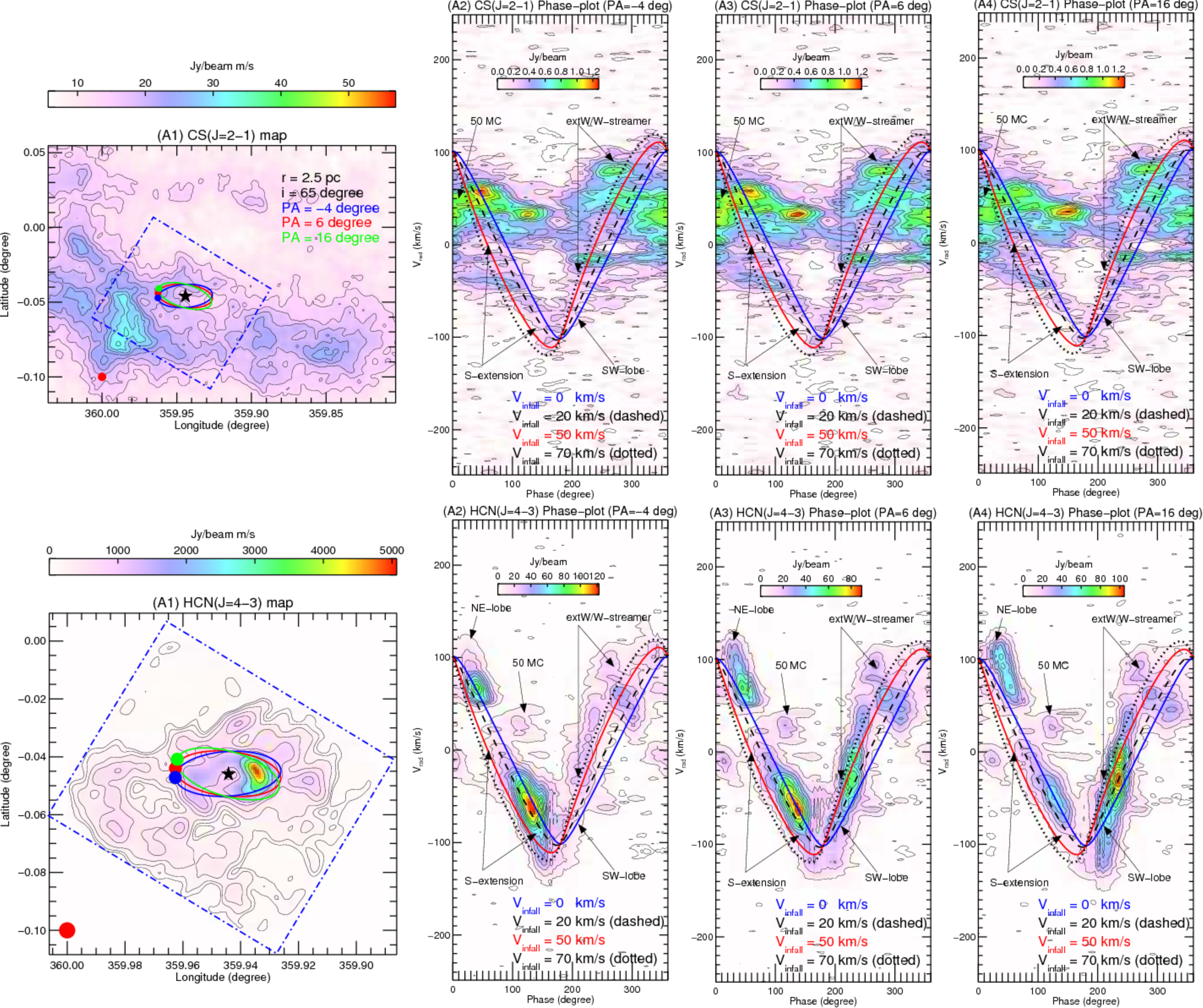}
\caption[]{Phase plots of the CND.
Upper: (A1) CS($J=2-1$) integrated intensity map. The beam size of 20$\arcsec$ is shown with the red dot in the lower left corner. The blue, red, and green elliptical annuli mark the trajectory of the phase plot (A2, A3, A4) with position angle of $-4\degr$, 6$\degr$, and 16$\degr$, respectively. The intrinsic radius is 2.5 pc. Starting position ($0\degr$) of the trajectory is labeled by the dots, and the phase is counterclockwise. The blue box marks the map size of the SMA data. Pure Keplerian ($V_{\rm infall}=0$ km s$^{-1}$; blue curve) and Keplerian plus infall models are overlaid on (A2-A4) with a radius of 2.5 pc, an inclination angle of 65$\degr$, and the PA of $-4\degr$, 6$\degr$, and 16$\degr$. The infall velocities are 20 km s$^{-1}$ (dashed line), 50 km s$^{-1}$ (red line) and 70 km s$^{-1}$ (dotted line), respectively. Lower: Same as above but for the HCN($J=4-3$) data.
}
\label{fig-phase-cnd2}
\end{center}
\end{figure}

\begin{center}
\begin{figure}
\epsscale{0.5}
\includegraphics[angle=0,scale=0.25]{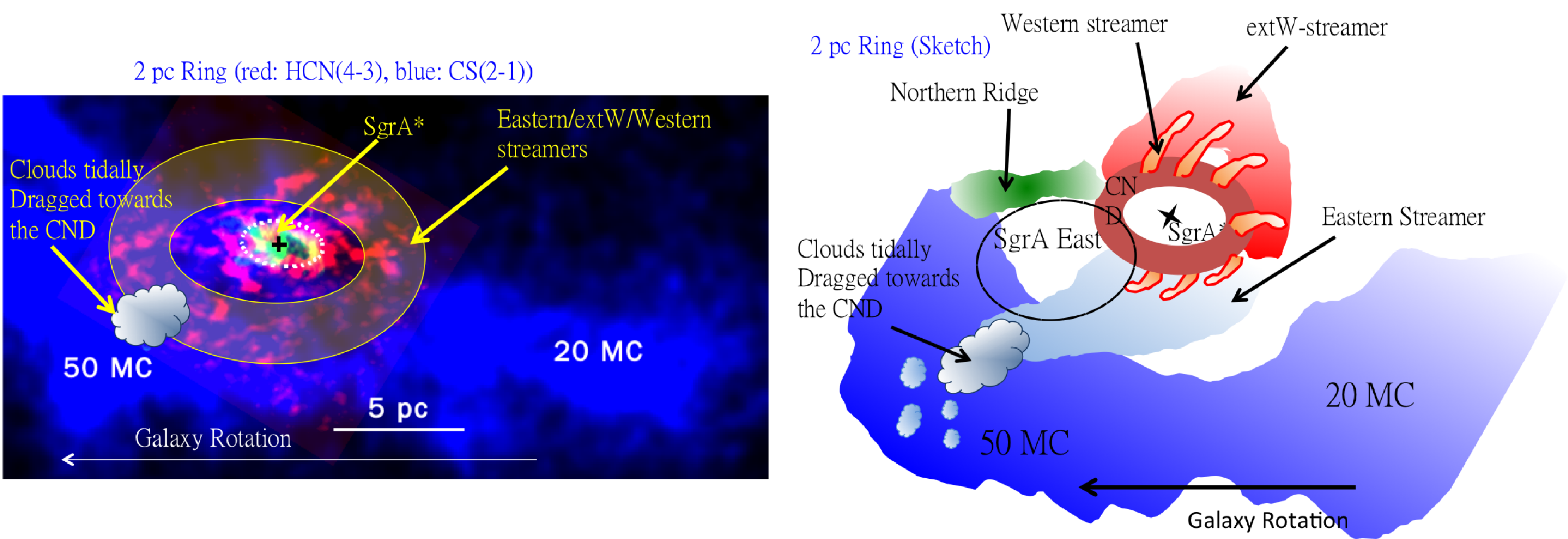}
\caption[]{Left: 3-color image. Blue color shows the CS($J=2-1$) data. Red color shows the SMA HCN($J=4-3$) data. Green color shows the VLA archival SgrA West in the 6-cm emission. Right: Schematic of the configuration of the 20/50 MC, the CND, SgrA East, the eastern streamer, and the extW/western-streamers. Our data show that the clumps are ripped away from the 50 MC and form the eastern streamers. The Eastern and western streamers can be described by a simple Keplerian rotation and infall motion. The western streamer observed in the interferometric HCN($J=4-3$) map is part of the CS($J=2-1$) ext-W streamer. The western streamer intersects with the CND as shown in the phase plots.}
\label{fig-model}
\end{figure}
\end{center}

\end{document}